\documentclass[10pt,conference]{IEEEtran}
\IEEEoverridecommandlockouts
\usepackage{cite}
\usepackage{amsmath,amssymb,amsfonts}
\usepackage{algorithmic}
\usepackage{graphicx}
\usepackage{textcomp}
\usepackage{xcolor}
\def\BibTeX{{\rm B\kern-.05em{\sc i\kern-.025em b}\kern-.08em
    T\kern-.1667em\lower.7ex\hbox{E}\kern-.125emX}}

\usepackage{xspace}
\newcommand{\tool}{\textsl{LiSum}\xspace}

\usepackage{cleveref}
\Crefformat{figure}{#2Fig.~#1#3}

\usepackage{comment}
\usepackage{pifont}
\usepackage{threeparttable}
\usepackage{amsmath}
\usepackage{verbatim}
\usepackage{bm}
\usepackage{graphicx}
\usepackage{subfig}
\usepackage{multirow}
\usepackage{enumitem}
\usepackage{xcolor}
\usepackage{booktabs}
\usepackage{makecell}

\newcommand{\lly}[1][]{#1}
\newcommand{\llyr}[1][]{#1}

\begin{document}

\title{LiSum: Open Source Software License Summarization with Multi-Task Learning}

\author{\IEEEauthorblockN{Linyu Li\IEEEauthorrefmark{1},
Sihan Xu\IEEEauthorrefmark{2}\IEEEauthorrefmark{3}\thanks{\IEEEauthorrefmark{3} Corresponding author. Email: xusihan@nankai.edu.cn},
Yang Liu\IEEEauthorrefmark{1},
Ya Gao\IEEEauthorrefmark{1},
Xiangrui Cai\IEEEauthorrefmark{1},
Jiarun Wu\IEEEauthorrefmark{1}, 
Wenli Song\IEEEauthorrefmark{4},
and Zheli Liu\IEEEauthorrefmark{2}}
\\
\IEEEauthorblockA{
\IEEEauthorrefmark{1}DISSec, NDST, College of Computer Science, Nankai University, Tianjin, China\\
\IEEEauthorrefmark{2}DISSec, NDST, College of Cyber Science, Nankai University, Tianjin, China\\
\IEEEauthorrefmark{4}Information Security Evaluation Center, Civil Aviation University of China, Tianjin, China\\
}
}

\maketitle

\begin{abstract}
Open source software (OSS) licenses regulate the conditions under which users can reuse, modify, and distribute the software legally. However, there exist various OSS licenses in the community, written in a formal language, which are typically long and complicated to understand. In this paper, we conducted a 661-participants online survey to investigate the perspectives and practices of developers towards OSS licenses. The user study revealed an indeed need for an automated tool to facilitate license understanding. Motivated by the user study and the fast growth of licenses in the community, we propose the first study towards automated license summarization. Specifically, we released the first high quality text summarization dataset and designed two tasks, i.e., license text summarization (LTS), aiming at generating a relatively short summary for an arbitrary license, and license term classification (LTC), focusing on the attitude inference towards a predefined set of key license terms (e.g., \textit{Distribute}). Aiming at the two tasks, we present \tool, a multi-task learning method to help developers overcome the obstacles of understanding OSS licenses. Comprehensive experiments demonstrated that the proposed jointly training objective boosted the performance on both tasks, surpassing state-of-the-art baselines with gains of at least 5 points \textit{w.r.t.} F1 scores of four summarization metrics and achieving 95.13\% micro average F1 score for classification simultaneously. We released all the datasets, the replication package, and the questionnaires for the community.

\end{abstract}


\section{Introduction}\label{sec:intro}

Open source software (OSS) has been widely-used in large-scale software systems to facilitate software development. Developers are permitted to incorporate the knowledge from OSS in their new implementations, so that they do not need to reinvent the wheel. Typically, an OSS is released under one or more licenses where the copyright holder grants licensees the rights to reuse, distribute, and modify the source code \textit{legally}. 
Anyone who reuses an OSS without conforming to the attached licenses may induce legal risks such as copyright infringement~\cite{chinacase,licensing2004software,infringement-case} and license violation~\cite{duan2017identifying}.

OSS licenses protect the rights for both OSS contributors and users. However, the regulations stated by OSS licenses are not easy and straightforward to understand. License texts, typically written in a formal language, are usually long and complicated, the average length of popular licenses from Choosealicense~\cite{url-choosealicense} is 1560 words. This leads to difficulty for many developers to comprehend. Moreover, there might also exist multiple licenses in one project, which further complicate matters~\cite{almeida2017software}. The number of official licenses continues to drastically grow, with over 552 listed by SPDX~\cite{url-spdx}. Furthermore, custom licenses with flexible expressions are becoming increasingly common, with a recent study of 1,846 projects identifying 5,777 unique licenses, 24.56\% of which were custom licenses \cite{xu2021lidetector}. Manually extracting the detailed information from these lengthy and complicated licenses is both time-consuming and labor-intensive. 
\iffalse
\begin{figure}
  \centering
  \includegraphics[width=0.98\linewidth]{sample/figures/license-text-summarization-tool-introduction-section.pdf}
  \caption{Two tasks for license understanding}
  \label{fig:intro}
\end{figure}
\fi

\begin{figure}
    \centering
    \includegraphics[width=0.95\linewidth]{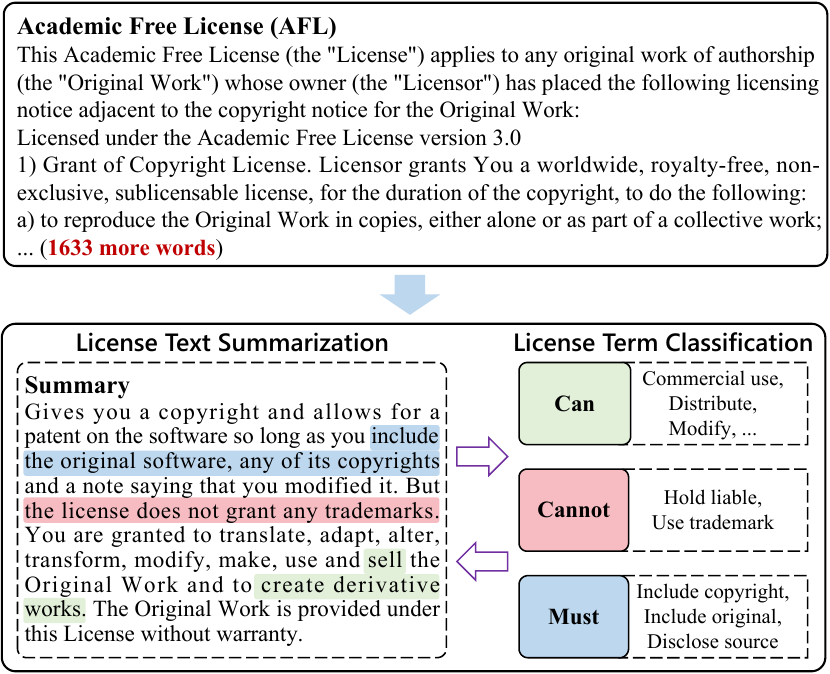}
    \caption{Two tasks for license understanding}
    \vspace{-7mm}
    \label{fig:intro}
\end{figure}

The aforementioned reasons indicate an indeed need to help OSS developers and users understand licenses, as it ensures compliance with legal obligations and avoids potential legal issues. 
To this end, much research has been done to model and extract the semantics of licenses. Majority of studies in the literature conducted the ontology study for license modeling~\cite{IREC`09-Alspaugh-Intellectual,AIS`10-Alspaugh-Challenge,Qualipso`10-gordon-prototype,ICAIL`11-Gordon-Analyzing,ICAIL`13-Gordon-Introducing,COMMA`14-Gordon-Demonstration}. For example, 
Gordon et al.~\cite{Qualipso`10-gordon-prototype} used the Web Ontology Language to build the ontology for 8 OSS licenses. Despite the efficiency, these works relied on expert experience to extract the tuples from license texts and were designed for a small set of licenses.
Unlike the ontology-based approaches that required substantial prior knowledge, Kapitsaki et al.~\cite{APSEC`17-Kapitsaki-termsIdentifying} presented the first tool, named FOSS-LTE, to automatically identify license regulations (e.g., MustOfferSourceCode) via a topic model. Specially, they first mapped licenses regulations with the phrases specified by manual analysis, and then mapped the phrases with the topics via Latent Dirichlet Allocation (LDA)~\cite{blei2003latent}. Finally, the topics were matched with predefined license regulations, by which they extracted the semantics of licenses. Despite the progress, these research on license understanding either relied on manual analysis or could only be applied for a predefined set of licenses, which limits their practical use. 

\llyr{Besides the research community, there is also a platform named TLDRLegal~\cite{url-tldrlegal}, where users can upload, summarize, and manage licenses. However, the quality of the summaries in TLDRLegal relies on the expertise of the users who uploaded them, leading to low-quality summaries in many cases. Specifically, the shortcomings of the summaries provided by TLDRLegal can be summarized from two aspects, i.e., incomplete information and requiring prior knowledge. (1) Incomplete information. For instance, the summary of the TOSG-2.0 license is ``A normal license''~\cite{url-tosg} which fails to mention the crucial aspects of the license, such as any liability disclaimers or warranty information. (2) Requiring prior knowledge. For example, the Academy of Motion Picture Arts and Sciences BSD License is summarized as ``A copyleft license, GPL incompatible''~\cite{url-academy-of-motion-picture-arts-and-sciences-bsd} which are not suitable for users who are unfamiliar with the GPL license. Another important problem of TLDRLegal comes from the quick adaption to new licenses or custom licenses. For instance, LLAMA2, a newly published model with significant importance, utilizes a custom license containing additional commercial terms, requiring licensees with over 700 million monthly active users to request additional grants from Meta~\cite{llama2-license}. Therefore, it is desirable to present a tool that can facilitate the understanding of an arbitrary license automatically.}

\llyr{In this paper, we present the first study towards automated license summarization. We observe that the characteristics of the dataset (i.e., pairs of the original license text and its summary) include three aspects. {First}, the available dataset was small-sized, which restricts the performance of deep learning models. {Second}, license text, usually written in formal language, is long and complicated, which is different from natural language text. {Third}, the summarization of OSS licenses have different focus than general summarization tasks; it pays more attention on the rights and obligations hidden in the license text. Due to these characteristics, we design \tool specialized for the domain of OSS licenses. Specifically, we propose a multi-task learning algorithm which leverages the correlation between two tasks (i.e., license text summarization and license term classification) to incorporate more information and make the model focus on the rights and obligations indicated by OSS licenses. As shown in \Cref{fig:intro}, the first task is license text summarization (LTS), aiming at generating a short and precise summary for an input license. The second task is license term classification (LTC), focusing on automatically inferring the attitudes conveyed by a license towards a set of key license terms (e.g., \textit{Commercial Use}). Since LTS and LTC are two highly correlated tasks, we propose a multi-task learning method to jointly learning two tasks and boost the performance of these tasks.} 

Comprehensive experiments demonstrated the effectiveness of the proposed jointly training objective. Specifically, \tool surpassed 7 state-of-the-art baselines in terms of Rouge-1/2/L and BERTScore, with gains of at least 5 points on all F1 scores for the LTS task. The comparative experiments also exhibited the superiority of \tool on the LTC task over existing baselines, with 95.13\% micro average F1 score. Moreover, the results of the ablation studies demonstrated the effectiveness of each module in \tool. 
We released all the datasets, the replication package, and the questionnaires for the community~\cite{url-LiSum}.

\noindent \textbf{Contributions}. In summary, we made the following novel contributions in this paper.
\begin{itemize}[leftmargin=*]
    \item We released the first high-quality dataset for license summarization, comprising 210 pairs of license full-texts and the summaries. To ensure the quality of our dataset, we conducted an online survey with 661 participants, and the feedback indicates that 76.04\% of the participants preferred the summaries in our dataset over TLDRLegal. 
    \item This study takes the first step towards automated license summarization. We proposed \tool, a multi-task learning method, to boost the performance on LTS and LTC simultaneously. Comprehensive experiments demonstrated that the proposed jointly training objective boosted the performance on both tasks simultaneously.
\end{itemize}
\section{Pre-trained Model}

{
Natural Language Processing is undergoing a paradigm shift due to the development of Pre-trained Language Models (PLMs).
Among these models are the masked language models such as BERT and RoBERTa ~\cite{kenton2019bert,liu2019roberta} have been shown to be highly effective for many downstream NLP tasks, such as text classification and entity extraction.
However, they often struggle with tasks that involve generating new text.
To address this limitation, the Bidirectional and Auto-Regressive Transformer (BART)~\cite{lewis2020bart} was proposed. BART is a sequence-to-sequence based denoising autoencoder that is pre-trained using a combination of denoising and generative tasks. Unlike masked language models, BART is specifically designed for text generation tasks, and has 
become one of the most commonly used models for text summarization. Therefore, in our study, we chose BART as the backbone of \tool.

\begin{figure}
    \centering
    \includegraphics[width=0.65\linewidth]{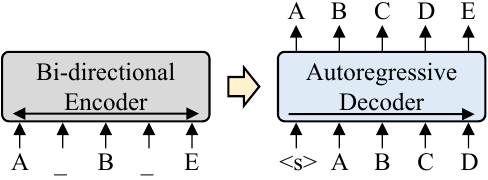}
    \vspace{-2mm}
    \caption{Overall architecture for training BART, where $<$s$>$ is a special token for shifting the sentence.}
    \vspace{-5mm}
    \label{fig:bart}
\end{figure}

The overall architecture of BART is depicted in \Cref{fig:bart}.
For training BART,
the corrupted input sentence (left, ``A \_ B \_ E'') is encoded with a bi-directional model (i.e., the Transformer model~\cite{vaswani2017attention}).
Then, the encoded vector is processed by an autoregressive decoder to reconstruct the original sentence (right, ``A B C D E''). 
During training, the teacher forcing strategy is utilized to train the decoder, i.e., the shifted output is used as the input of the encoder.
Since BART employs an autoregressive decoder, it can be directly fine-tuned for text generation tasks, such as text summarization.
To fine-tune a summarization model based on BART, the input of the encoder is the original text, and the decoder learns to generate the responding text summary autoregressively.
\section{Approach}
\tool is an end-to-end multi-task network that tackles both license text summarization and license term classification. This section describes our approach in detail.

\subsection{Problem Statement}
Given an OSS license $D$ with $l$ words $D=(w_1, w_2, \dots, w_l)$, the license understanding problem in this paper contains two sub-tasks.
The first task is to achieve an appropriate summary $V=(v_1, v_2, \dots, v_k)$ with $k$ words.
The second task is to infer the attitudes towards a predefined set of licence terms based on the original licence text.
Following previous studies~\cite{url-tldrlegal,xu2021lidetector}, this paper focuses on inferring the attitudes of 23 key license terms.
We list these license terms in \Cref{tab:terms}, and the detailed descriptions of them can be seen online~\cite{url-LiSum}.
We denote the license terms by $\{t_1, t_2, \dots, t_{23}\}$.
For each license term, there are four different attitudes according to the license, i.e., \texttt{can}, \texttt{cannot}, \texttt{must}, and \texttt{not mentioned}.
Therefore, we formulate the attitude inference problem of the 23 license terms as 23 classification problems.
For the $i$-th term, the attitude inference achieves a predicted label $\hat y_i = \phi(D)$, where $\hat y_i \in \{\texttt{can, cannot, must, not mentioned}\}$ and $\phi(\cdot)$ is a classification model.

\begin{table}[]
    \vspace{-2mm}
    \caption{List of License Terms}
    \vspace{-1mm}
    \centering
    \footnotesize
    \begin{tabular}{ccc}
    \toprule
    \textbf{Line NO.} & \textbf{License Terms}     &  \textbf{License Terms} \\
    \midrule
     1 & Distribute    & Modify \\
     2 & Commercial Use & Hold Liable\\
     3 & Include Copyright & Include License \\
     4 & Sublicense & Use Trademark \\
     5 & Private Use & Disclose Source \\
     6 & State Changes & Place Warranty \\
     7 & Include Notice & Include Original \\
     8 & Give Credit & Use Patent Claims \\
     9 & Rename & Relicense \\
     10 & Contact Author & Include Install Instructions \\
     11 & Compensate for Damages & Statically Link \\
     12 & Pay Above Use Threshold &\\
     \bottomrule     
    \end{tabular}
    \vspace{-5mm}
    \label{tab:terms}
\end{table}

\subsection{Model Architecture}

\iffalse
\begin{figure*}[h]
  \centering
  \includegraphics[width=0.78\linewidth]{sample/figures/model.pdf}
  \caption{Overview of \tool, a multi-task learning model that consists of a summarization module and a classification module. We employ BART as the backbone, and propose a multi-task objective to jointly train two sub-tasks. In the classification module, we design a term attention module to learn a term-specific representation for each license term.}
  \label{fig:model_overview}
\end{figure*}
\fi

\begin{figure*}[h]
  \centering
  \includegraphics[width=0.85\linewidth]{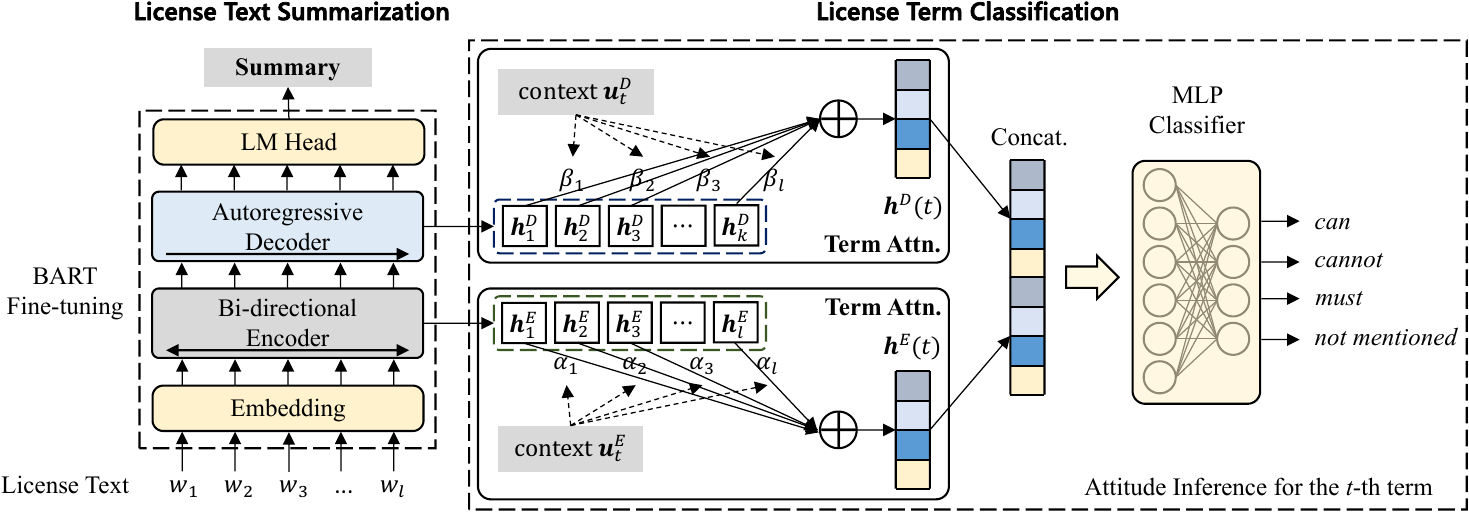}
  \vspace{-1mm}
  \caption{Overview of \tool, a multi-task learning model that consists of a summarization module and a classification module. We employ BART as the backbone, and propose a multi-task objective to jointly train two sub-tasks. In the classification module, we design a term attention module to learn a term-specific representation for each license term.}
  \label{fig:model_overview}
  \vspace{-5mm}
\end{figure*}

As illustrated in \Cref{fig:intro}, there are high correlations between a high-quality license summary and the license terms related to the regulations, obligations, and conditions of software use. Inspired by this observation, we propose a multi-task learning model (i.e. \tool) to incorporate more information and make the summary concentrate on the regulations required by licenses. By jointly learning both tasks through multi-task learning, we expect to enhance the understanding of licenses and improve the performance of both tasks simultaneously.

\Cref{fig:model_overview} depicts the overall architecture of \tool. Specifically, \tool consists of a license summarization module and an attitude classification module. In this architecture, the bi-directional encoder processes the input license text and generates contextualized representations at each token position. The autoregressive decoder is responsible for the text summarization task. Moreover, we introduce a novel term attention module designed specifically for attitude inference. This module enhances the model to focus on specific license terms during the classification process. Consequently, the model can effectively extract relevant information for each term, even from lengthy and complex licenses, while keeping the number of additional parameters to a minimum. The license representations from the encoder and decoder are fed into the term attention module to acquire term-specific representations, which are then concatenated to create a fused representation. This fusion aims to capture complementary information, effectively enriching the understanding of license terms. The concatenated representation serves as the foundation for the subsequent classification task, which is performed by a linear classifier.
Finally, it is worth noting that while we utilize BART as the backbone, \tool can also be constructed based on other sequence-to-sequence pre-trained language models.

\subsection{Licence Text Summarization}

The goal of the license text summarization module is to generate a summary with a few words for a given license, so as to facilitate license understanding.
Generally, text summarization aims to compress a document into a brief summary while keeping the salient information in the original document.
There are mainly two types of approaches for text summarization, i.e., extractive and abstractive text summarization.
The extractive methods~\cite{zhong2020extractive,narayan2020stepwise} select salient sentences or words from the original text and concatenate them as the final summary.
The abstractive methods~\cite{zhang2020pegasus,maynez2020faithfulness} generate new words or sentences, and then rephrase them to summarize the input text, which are usually more flexible and human-like.

Considering the characteristics of OSS license summaries, we propose an abstractive approach for license text summarization in this paper.
Intuitively, an OSS license, as a form of legal document, has its own characteristics, which is different from news articles~\cite{kanapala2019text}.
First, an OSS license is usually long.
Summarization of an OSS license requires a high compression rate relative to the original license.
Second, OSS license contains some obscure jargon (e.g., ``sublicense'', ``relicense'') and the salient information is scattered throughout the whole license. Third, license is a low-resourced language without a large corpus to train.
Due to these characteristics, it is difficult for the extractive methods to obtain accurate license summaries.
To verify the hypothesis, we followed the approach proposed by Fabbri et al.~\cite{fabbri2020improving}. Specifically, they binned the abstractive level of summaries into four categories, ranging from extremely extractive to extremely abstractive by calculating the ROUGE score~\cite{lin2004rouge} of summaries and source documents. Our results were consistent with the intuition that license summaries tend to be more abstractive rather than extractive, as indicated by %The results are as follows: ROUGE-1 F1: 17.19\%, ROUGE-2 F1: 11.04\%, ROUGE-L F1: 21.20\%. As we can observe, 
the ROUGE scores being lower than 30\%, which is the boundary between more abstractive and more extractive summaries.

In this paper, we employ BART, a sequence-to-sequence pre-trained language model, as the backbone of \tool.
%Recently, abstractive summarization is dominated by Transformer-based models~\cite{vaswani2017attention}.
%As a sequence-to-sequence pre-trained language model, BART is suitable for text generation tasks. 
%Note that we can also build the model based on other sequence-to-sequence pretrained language models.
Specifically, we take the OSS license text $D=(w_1, w_2, \dots, w_l)$ as the input of BART.
The model first obtains the embeddings of the words, i.e., $(\bm x_1, \bm x_2, \dots, \bm x_l) = E(w_1, w_2, \dots, w_l)$, where $E(\cdot)$ is the embedding layer.
Then, we achieve the hidden representations of the license by the bi-directional encoder, formally, 
\begin{equation}
    (\bm h_1^E, \bm h_2^E, \dots, \bm h_l^E)=\mathcal F(\bm x_1, \bm x_2, \dots, \bm x_l) \,,
\end{equation}
where $\mathcal F$ refers to the bi-directional encoder.
The encoder is built with multiple Transformer layers (i.e., 6 layers for BART-base and 12 layers for BART-large).
{
The encoded representations $(\bm h_1^E, \bm h_2^E, \dots, \bm h_l^E)$ are then decoded by the autoregressive decoder, i.e.,
\begin{equation}
    (\bm h_1^D, \bm h_2^D, \dots, \bm h_k^D)=\mathcal{G}(\bm h_1^E, \bm h_2^E, \dots, \bm h_l^E) \,,
\end{equation}
where $\mathcal G$ refers to the decoder, and $k$ the length of the summary.
The autoregressive decoder is built with the same number of Transformer layers as that of the encoder.
We use the teacher forcing strategy for training the autoregressive decoder.
During training, the shifted summary is utilized as the input of the decoder, while the output vectors of the encoder are the input during inference.
}
{
After that, the sequence of hidden vectors $(\bm h_1^D, \bm h_2^D, \dots, \bm h_k^D)$ is fed to a Language Model Head (i.e., LM Head) to obtain the license summary, where LM head is a linear layer having input dimension of hidden vectors and output dimension of vocabulary size, i.e.,
\begin{equation}
    \hat{\bm v}=\text{Linear}(\bm h_1^D, \bm h_2^D, \dots, \bm h_k^D)\,,
\end{equation}
where $\hat{\bm v}\in \mathbb R^{S}$ is the output summary and $S$ refers to the vocabulary size.
Finally, we minimize the cross entropy between the generated summary $\hat{\bm v}$ and the ground truth $\bm v$.
\begin{equation}
    \mathcal{L}_s=-\sum_{i=1}^S v_i \log \hat v_i \,.
    \label{eq:loss_sum}
\end{equation}
}

{
Since license text is a low-resourced language, to avoid overfitting, we adopt a two-stage fine-tuning strategy, which has been widely used in abstractive summarization on low resources~\cite{su2021improve,bajaj2021long}.
Specifically, we first fine-tune BART on the CNN dataset~\cite{hermann2015teaching}, equipping the model with the ability of abstract extraction. 
Then, we fine-tune the model on our dataset, so that the model can adapt to the abstractive summarization task for OSS licenses.
The ablation study in the Supplementary Material demonstrated the benefits of the two-stage fine-tuning implemented in \tool. %(Please refer to \Cref{sec:RQ4} for the details.)
}

\subsection{License Term Classification}\label{sec:classfication}

The design of our license term classification module stems from two primary considerations, both of which are deeply rooted in the essence of license understanding. First, valuable feedback from our user study (\Cref{sec:userstudy}) suggested that presenting structural information with highlighted attitudes could significantly improve license understanding. In response to these concerns, we developed the license term classification module to automatically infer attitudes toward key license terms (e.g., \textit{Distribute}), thereby highlighting crucial aspects of the licenses and facilitating users' comprehension. Second, from a technical standpoint, the classification module plays a pivotal role in our multi-task learning approach. As mentioned earlier, license term classification and license text summarization share inherent correlations. By simultaneously learning both tasks, our model (i.e., \tool) leverages their interplay to enhance the overall license understanding process and generate more informative and concise license summaries.

Like previous studies{~\cite{url-tldrlegal,xu2021lidetector}}, we focused on the attitudes conveyed by licenses towards a predefined set of key license terms as shown in \Cref{tab:terms}.
%We model the attitude inference as classification tasks.
For each license term, the input is an original license text $D=(w_1, w_2, \dots, w_l)$, and the output label $y$ is one of the four possible attitudes, i.e., \texttt{can}, \texttt{cannot}, \texttt{must}, and \texttt{not mentioned}.
Therefore, the attitude inference problem is formulated as a multi-class classification task.
One straightforward method for license term classification is to build 23 classifiers individually, where each model classifies the attitudes towards a single license term.
However, such 23 classifiers are parameter-redundant since the license representation learning can be shared across different classifiers.
Therefore, we propose a novel term attention module to disentangle license representations for different license term classifiers.
Compared with individual modelling, training the 23 classification tasks with a uniform model is beneficial to learning license representations.

We first extract a license representation from the BART backbone.
Specifically, we obtain license representations by the output of the bi-directional encoder and that of the autoregressive decoder, i.e., $(\bm h_1^E, \bm h_2^E, \dots, \bm h_l^E)$ and $(\bm h_1^D, \bm h_2^D, \dots, \bm h_k^D)$.
Then, we design a term attention module that learns to attend different hidden states according to different license terms.
For the $t$-th license term, we can obtain the attention weights by
\begin{equation}
    \alpha_i = \frac{\exp{(\bm h_i^E \cdot \bm u_t^E)}}{\sum_{m=1}^l \exp{(\bm h_m^E \cdot \bm u_t^E)}} \,,
    \label{eq:attention}
\end{equation}
where $\alpha_i$ is the attention weight on $\bm h_i^E$, $i \in \{1, 2, \dots, l\}$, and $\bm u_t^E$ is a learnable context vector corresponding to the $t$-th license term.
Then, we can obtain a term-specific license representation from the encoder by the weighted average of the hidden vectors, i.e.,
\begin{equation}
    \bm h^E(t) = \sum_{i=1}^l \alpha_i \bm h_i^E \,. \label{eq:weighted_sum}
\end{equation}

{
To enhance license representation, \tool also incorporates the output of the decoder.
Similar to Equation~\eqref{eq:attention} and~\eqref{eq:weighted_sum}, we perform the term attention to aggregate the information from the decoder. Formally, 
\begin{equation}
    \begin{aligned}
    &\beta_j = \frac{\exp{(\bm h_j^D \cdot \bm u_t^D)}}{\sum_{m=1}^k \exp{(\bm h_m^D \cdot \bm u_t^D)}} \,,~ 
    &\bm h^D(t) = \sum_{j=1}^l \beta_j \bm h_j^D \,,
    \end{aligned}
\end{equation}
where $\beta_j$ is the attention weight on $\bm h_j^D$, $j \in \{1, 2, \dots, k\}$, $\bm u_t^D$ is a learnable context vector corresponding to the $t$-th license term.
We obtain the term-specific representation from the decoder perspective by the weighted average.
We initialize the context vectors $\bm u_t^E$ and $\bm u_t^D$, $t\in \{1, 2, \dots, 23\}$ with the semantics of license terms. Specifically, for a license term $\text{term}_t$, we feed it into the pre-trained BART encoder and decoder, and then initialize $\bm u_t^E$ and $\bm u_t^D$ with the outputs of the encoder and decoder, respectively. 
\iffalse
\begin{itemize}
    \item \textit{Random Initialization}, which initializes the context vectors randomly, e.g., with Gaussian distribution.
    \item \textit{Term Semantic Injection}, where we initialize the context vectors with pretrained license term vectors.
    For instance, we can initialize $\bm u_t^E$ by the output of the BART encoder, $\bm u_t^E=\mathcal{F}(\text{term}_t)$, where $\text{term}_t$ is the word or phrase of the $t$-th term, and $\mathcal{F}$ the pretrained BART encoder.
\end{itemize}
\fi
}

{
We concatenate the two term-specific representations, i.e., $\bm h(t) = \bm h^E(t) || \bm h^D(t)$, to achieve the final term-specific representation, and develop a multi-layer perception as the classifier.
First, we transform $\bm h(t)$ with a linear layer, i.e.,
\begin{equation}
    \bm h^\prime(t) = \text{ReLU}(\bm W_1 \bm h(t) + \bm b_1) \,,
\end{equation}
where $\bm W_1$ and $\bm b_1$ are the weight and bias parameters within the linear layer, and ReLU the activation function.
We utilize the dropout technique~\cite{srivastava2014dropout} to avoid overfitting, i.e., $\bm h^{\prime\prime}(t)=\text{Dropout}(\bm h^\prime(t))$.
Finally, we obtain the predicted class of the $t$-th license term through another linear layer, i.e.,
\begin{equation}
    \hat{\bm y}(t) = \bm W_2 \bm h^{\prime\prime}(t) + \bm b_2 \,,
\end{equation}
where $\bm W_2$ and $\bm b_2$ are the weight and bias parameters within the linear layer, and $\bm y(t) \in \mathbb{R}^4$.
}

{Finally, we utilize cross entropy as the loss function.
Given an input license, the loss of license term classification is defined by
\begin{equation}
    \mathcal{L}_c = -\frac{1}{23}\sum_{t=1}^{23}\sum_{i=1}^4 y(t)_i \log \hat y(t)_i \,,
    \label{eq:loss_cls}
\end{equation}
where $y(t)_i, i\in \{1,2,3,4\}$ denotes the ground truth.
Note that \tool can be trained for the 23 license term classification tasks together through the term attention module.
}

\subsection{Multi-task Objective}
{
As described in \Cref{sec:intro}, license text summarization and license term classification are highly correlated tasks.
Therefore, we combine them by sharing the backbone model and jointly training the two objectives in Equation~\eqref{eq:loss_sum} and~\eqref{eq:loss_cls}.
Given an input license text, the multi-task objective of \tool is defined as follows:
\begin{equation}
    \mathcal L = \mathcal{L}_s + \lambda \mathcal{L}_c \,,
\end{equation}
where $\lambda$ is a hyper-parameter to trade off the weights of the two tasks.
We use the AdamW algorithm~\cite{loshchilov2019decoupled} to minimize this loss function and train the parameters.
}
\section{Experiments}
In this section, we conduct an online user study and an experimental study to answer the following research questions:

\textbf{RQ1:} Do OSS developers need license summaries to facilitate the understanding of OSS licenses?

\textbf{RQ2:} How effective is \tool for summarizing license text compared with state-of-the-art text summarization baselines?

\textbf{RQ3:} Can \tool outperform the state-of-the-art techniques in inferring the attitudes conveyed by licenses? 

% \textbf{RQ4:} How do the modules in \tool effect the performance on license text summarization and license term classification?

%\textbf{RQ4:} What are the feedbacks from OSS developers for \tool?

\subsection{The Referenced Dataset} \label{sec:dataset}

To facilitate the understanding of OSS licenses, two platforms (i.e., {Choosealicense}~\cite{url-choosealicense} and {TLDRLegal}~\cite{url-tldrlegal}) provide manually crafted license summaries. Specifically, Choosealicense is a website where developers can find summaries and full-texts of some popular licenses. Despite the usefulness, there are only 45 licenses supported by Choosealicense and TLDRLegal provides some additional licenses.  Although these two platforms cover the majority of common licenses, the number of licenses increases significantly each year, and custom licenses are also numerous, making users having difficulty in understanding other uncommon licenses. Moreover, TLDRLegal allows individuals to upload licenses and summaries and the quality of these summaries varies greatly and depends on the individual's understanding and expertise of the licenses. \llyr{For instance, the summaries of the TOSG-2.0 license~\cite{url-tosg} provides incomplete information and the LLGPL license~\cite{url-llgpl} suffers from the issues of requiring prior knowledge, These issues contribute to the overall poor quality of the summaries provided by TLDRLegal.}

Since there are various official licenses, variants, custom licenses, and the number of licenses keeps growing, we propose to construct a machine-learning model for automated license text summarization. However, there existed no ground-truth dataset of high quality for training and evaluation. To address this issue, we first construct a high-quality dataset of license summarization. \llyr{Specifically, we first selected 45 common licenses from Choosealicense~\cite{url-choosealicense} and 165 licenses from TLDRLegal~\cite{url-tldrlegal}, respectively. These licenses cover all 23 license terms. Then, four authors independently produced summaries for all licenses according to the following criteria. (1) Correctness. A license summary is supposed to be consistent with the content of original license text. (2) Necessary information. A license summary should cover all regulations required by the original license to avoid license misuse and violation. (3) Readability. A high-quality license summary should be readable to all users without requiring any prior knowledge or obscure statements. (4) Conciseness. To facilitate license understanding, a license summary is supposed to be concise, without statements irrelevant to rights, obligations, and conditions of software use. Finally, we cross-validated all license summaries and voted for a best summary for each license to construct a high-quality dataset for license summarization.} In total, we obtained a dataset comprising 210 pairs of license full-texts and the summaries and made the license summary dataset publicly available online~\cite{url-LiSum}. We also performed a user study to evaluate the quality and effectiveness of the referenced dataset, which can be seen in \Cref{sec:RQ1}.

\subsection{RQ1: User Study}\label{sec:RQ1} \label{sec:userstudy}
To study the necessity of license understanding and investigate the quality of the license summary dataset, we conducted an online user study in both English and Chinese. The contents of two questionnaires are consistent across languages. We made the questionnaires and answers publicly online~\cite{url-LiSum}.

\begin{table}
\vspace{-2mm}
\caption{Demographics of the questionnaire participants.}
\label{tab:demographics}
\footnotesize
\centering
\begin{tabular}{cccccc} 
\toprule
\multirow{2}{*}{\textbf{Demographics}}                    & \multirow{2}{*}{\textbf{Groups}} & \multicolumn{2}{c}{\textbf{Prolific}}                              & \multicolumn{2}{c}{\textbf{Wenjuanxing}}                            \\
                                                 &                         & NO. & Percentage & NO. & Percentage \\ \midrule
\multirow{4}{*}{Age}                             & $\textless$18                      & 0                          & ~0.0                                 & 3                          & ~1.1                                  \\
                                                 & 18-25                   & 95                         & ~37.5                                & 196                        & ~74.5                                 \\
                                                 & 26-40                   & 111                        & ~43.8                                & 55                         & ~20.9                                 \\
                                                 & $\textgreater$40                      & 47                         & ~18.7                                & 9                          & ~3.4    \\\midrule                             
\multirow{4}{*}{Gender}                          & Female                  & 184                        & ~72.7                                & 173                        & ~65.8                                 \\
                                                 & Male                    & 68                         & ~26.9                                & 80                         & ~30.4                                 \\
                                                 & No-binary               & 1                          & ~0.4                                 & 1                          & ~0.4                                  \\
                                                 & No answer               & 0                          & ~0.0                                 & 9                          & ~3.4                                  \\ \midrule
\multirow{6}{*}{\makecell{\lly{Programming}\\ Experience}}                 & no exper.                      & 6                          & ~2.3                                 & 11                         & ~4.2                                  \\
                                                 & $\textless$1 year                  & 47                         & ~18.6                                & 17                         & ~6.5                                  \\
                                                 & 1-3 years               & 73                         & ~28.9                                & 61                         & ~23.2                                 \\
                                                 & 3-5 years               & 51                         & ~20.2                                & 96                         & ~36.5                                 \\
                                                 & 5-10 years              & 36                         & ~14.2                                & 60                         & ~22.8                                 \\
                                                 & $\textgreater$10 years                & 40                         & ~15.8                                & 18                         & ~6.8               \\\midrule
Total & / & 253 & 100 & 263 & 100 
                                                 \\\bottomrule

\end{tabular}
\vspace{-2mm}
\end{table}

\noindent \textbf{Recruitment and demographics}. We recruited 661 participants via two online research platforms (i.e., 339 in English by Prolific~{\cite{url-prolific}} and 322 in Chinese by Wenjuanxing~{\cite{url-wjx}}). 
We started the survey on December 2022 and spent one month to collect the responses.  
\llyr{To conduct an effective study, we select participants who at least graduated from high school and have experience in programming and OSS. Specifically, for the requirement of the education background and programming skills, we leveraged the platform's automated filtering mechanism to screen the participants based on the information provided in their profiles. For the requirement of the experience in the OSS community, we added a description on the top of the questionnaire. To validate the background of participants, we also include questions about the education background, programming skills, and OSS experience in the questionnaire. Moreover, we utilized a systematic approach to filter out possible low-quality responses. Specifically, we calculated the 10th percentile of the time taken by participants to complete their questionnaires, which was 122.2 seconds. It indicates that only 10\% of the participants took less than 122.2 seconds to complete the survey. For this reason, we set a 2-minute threshold to remove samples that deviate significantly from the normal samples.} In total, we obtained 253 responses from Prolific and 263 responses from Wenjuanxing. The average time of completing the survey is 6 minutes.
According to the local income, we payed each participant in Wenjuanxing 6 CNY and payed each participant in Prolific 1.5 GBP.
Owing to the prescreened demographics, all participants have graduated from high school; 96.71\% of the participants have programming skills; 85.08\% of the participants have experiences in the OSS community. The demographics of the participants can be seen in \Cref{tab:demographics}. We have released the contents of the questionnaires and the responses~\cite{url-LiSum}.

\begin{figure}[t]
  \centering
  \includegraphics[width=0.8\linewidth]{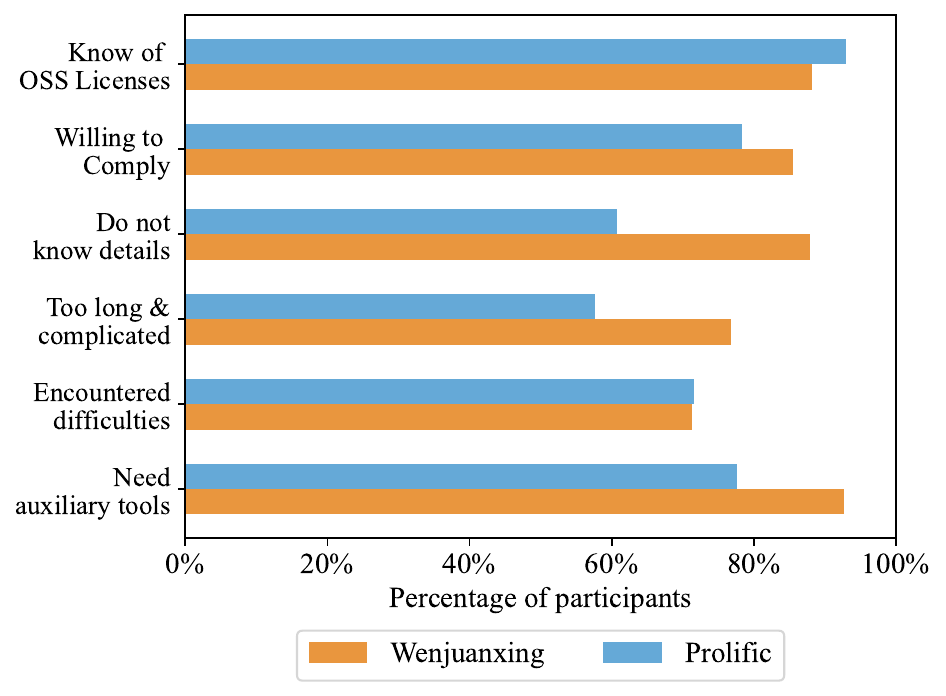}
  \vspace{-2mm}
  \caption{Perspectives and practices towards OSS licenses.}
  % \Description{Participants from Prolific and Wenjuanxing have similar feelings of OSS licenses}
  \vspace{-6mm}
  \label{fig:difficulty}
\end{figure}

\noindent \textbf{Perspectives and practices towards OSS licenses}. We first collected the perspectives and practices of the participants towards OSS licenses. 
{\Cref{fig:difficulty} displays the attitudes of the participants towards OSS licenses. }It can be seen that the participants in Prolific (with English questionnaires) and those in Wenjuanxing (with Chinese questionnaires) have similar feelings of OSS licenses. Specifically, in total, 90.50\% of the participants have heard about OSS licenses (Q7/8). 81.98\% of the participants would like to comply with the OSS licenses when they reused open source code (Q13). Out of all the participants, 57.75\% reported having experienced in choosing OSS licenses for their own implementations (Q12). Among these participants, a substantial proportion (71.43\%) have encountered difficulty when making such decisions (Q11). Finally, the results of Q14 and Q15 show that 69.29\% of the participants thought OSS licenses were too long and complicated to understand, and 80.05\% of the participants reflected that they needed an automated tool to facilitate the understanding of OSS licenses.

For statistical analysis, we exploited the Chi-squared test to estimate the relations between the variables in the survey. Specifically, we first measure the relations between the experience in reusing OSS code (Q5) and the awareness of OSS licenses (Q13), and the relations between the experience in developing OSS (Q6) and the practice in choosing OSS licenses (Q12). According to the analysis of the Chi-squared test, we found that OSS users tend to have the willing to comply with the licenses of the OSS ($\mathcal{X}^2(2)=21.412, p<0.001$). Similarly, OSS developers tend to choose an appropriate OSS licenses for their own projects  ($\mathcal{X}^2(2)=47.651, p<0.001$). Despite the awareness of intellectual property protection, the Chi-squared test proved that OSS developers who needed OSS licenses tend to have the difficulty in choosing OSS licenses ($\mathcal{X}^2(1)=26.009, p<0.001$). Finally, we found the correlation between the difficulty in choosing OSS licenses (Q11) and the difficulty in understanding OSS licenses (Q14) ($\mathcal{X}^2(1)=4.711, p=0.030$). To sum up, according to the quantitative analysis of the Chi-squared test, it can be inferred that although most OSS developers claimed they were aware of OSS licenses, they have difficulty in using OSS licenses in their practice, which are probably due to the long and complicated license text. Therefore, it is desirable to present a tool to facilitate the understanding of OSS licenses, so as to promote the practical use of OSS licenses. The full results of the survey can be seen in~\cite{url-LiSum}.

\begin{figure}[t]
\vspace{-3mm}
    \centering
    \subfloat[Wenjuanxing]{
    \includegraphics[width=0.24\textwidth]{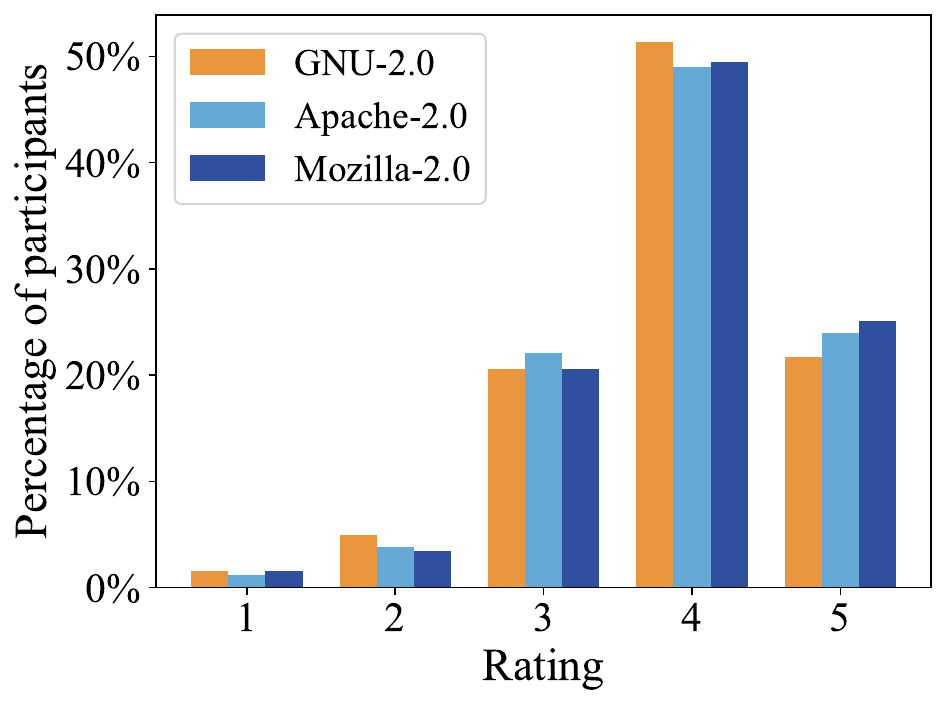}
    \label{Fig:cvebar}}    
    %\quad
    \subfloat[Prolific]{
    \includegraphics[width=0.24\textwidth]{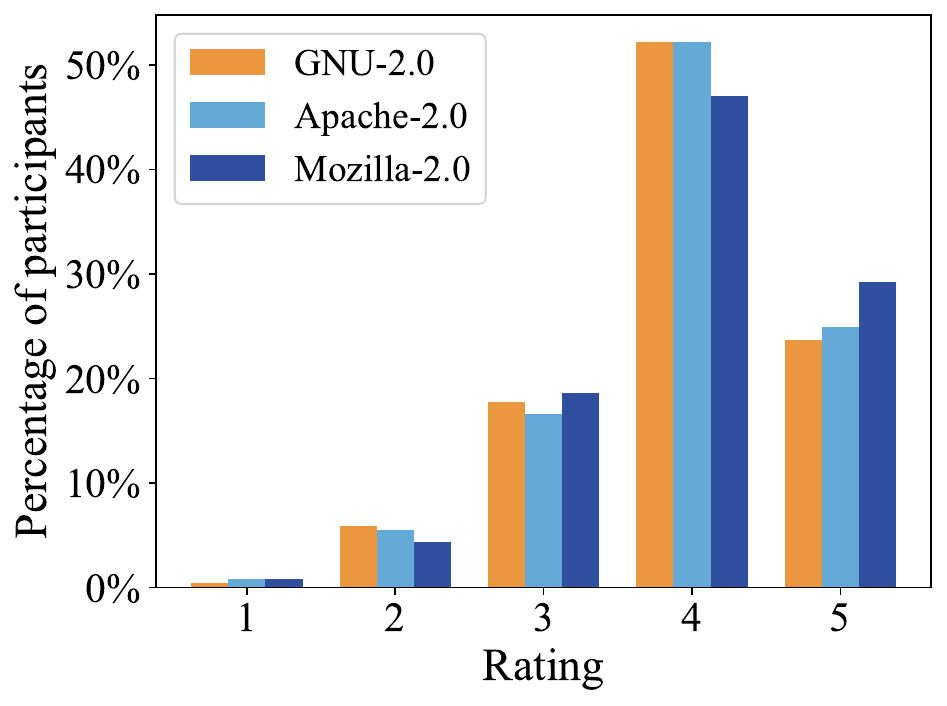}
    \label{Fig:mthbar}
    }  
    \vspace{-2mm}
    \caption{Scores of the summaries of three popular licenses.}
    \vspace{-6mm}
    \label{fig:augmenttypes}
\end{figure}

\noindent \textbf{Quality of the license summary dataset}. We also collected the attitudes of the participants towards license summary, especially the labelled license summary dataset. We first took three popular licenses (i.e., GPL-2.0, Apache-2.0, and MPL-2.0) as the examples, and asked the participants whether the summaries in our dataset contribute to a quick understanding of the regulations stated by these licenses (Q16-Q18). The participants gave each summary a score from 1 to 5 (i.e., 1 denotes ``not helpful'' and 5 denotes ``very helpful''). \Cref{fig:augmenttypes} shows the scores of the labelled summaries of three popular licenses. In total, the average scores of GPL-2.0, Apache-2.0, and MPL-2.0 are {3.90, 3.93, and 3.96}, respectively. It can be inferred that the participants tend to think the license summaries were helpful for them to understand these popular licenses.

\iffalse
\begin{figure}[h]
  \centering
  \subfigure[Wenjuanxing]{
      \includegraphics[scale=0.4]{sample/figures/ratings_of_summaries_wenjuanxing.pdf}
      \Description{Scores of summaries rated by participants from Wenjuanxing}
  }
  \hspace{0.5in}
  \subfigure[Prolific] {
      \includegraphics[scale=0.4]{sample/figures/ratings_of_summaries_prolific.pdf}
      \Description{Scores of summaries rated by participants from Prolific}
  }
  \caption{Scores of 3 popular licenses' summaries}
  
\end{figure}
\fi

We also compared the labelled license summaries with those provided by TLDRLegal~\cite{url-tldrlegal}. To conduct a fair comparison, each participant was assigned 5 pairs of license summaries randomly selected from the 210 pairs of license summaries. In total, 516 participants answered 2580 questions in this part. 76.04\% of these answers reflected that the participants preferred the license summary we labelled compared to the one provided by TLDRLegal. By analyzing the results, we summarized several reasons why some participants preferred the license summaries in TLDRLegal. \ding{172} In some cases, license summaries in TLDRLegal are more \lly{vague} than those in our dataset. For instance, the summaries of {the Haxcv License} are ``\textit{Basically, you can do whatever you want...}'' in TLDRLegal and ``\textit{You can use, copy, modify...}'' in the labelled dataset. Although the former is easier to understand, the labelled dataset can provide users a more precise license summary. The reason is that even in the permissive licenses, not all rights are explicitly granted. \ding{173} Some licenses in TLDRLegal required prior knowledge to understand them. For example, in {the BitTorrent Open Source License v1.1}, the summary in TLDRLegal is ``\textit{Similar to the Mozilla Public License, this license...}''. Despite that some participants might prefer this type of license summary, we did not required any prior knowledge from users in our dataset. \ding{174} Some license summaries are very similar in two datasets with few different words ({e.g., Foxit EULA, FusionLord Custom License}). In this case, the participants might randomly select an answer.

\noindent \textbf{Suggestions}. At the end of the questionnaire, there was an optional question (Q24) that asked for any suggestions toward license summaries. 29 out of 120 suggestions mentioned that structural information (e.g., a list or a table of license terms) might benefit understanding licenses at a glance. 20 participants suggested shorter and simpler summaries to assist them in understanding licenses. 3 participants preferred license summaries with highlighted attitudes (e.g., \textit{Can}). Moreover, there were also a few suggestions that mentioned the definitions of some license terms, the examples, and the emphasis on certain rights or obligations in the summaries.

\noindent \fbox{
	\parbox{0.95\linewidth}{
	\textbf{Answer to RQ1:} %The user study revealed that it is desirable to mitigate the obstacles of understanding OSS licenses. 
 71.43\% of the participants have encountered difficulties when choosing OSS licenses. 69.29\% of the participants thought OSS licenses were too long and complicated. 80.05\% of the participants reflected that they needed an automated tool to facilitate license understanding. Moreover, 76.04\% of the feedbacks reflected the preference of the summaries we labelled compared with those in TLDRLegal.}
}

\begin{table*}[]
\small
    \caption{Recall (R), precision (P), and F1 score of license text summarization (\%)}
    \vspace{-1mm}
    \label{expe:summary}
    \centering
    \begin{tabular}{ccccccccccccc}
    \toprule
    \multirow{2}{*}{\textbf{Model}}& \multicolumn{3}{c}{\textbf{Rouge-1}}                         & \multicolumn{3}{c}{\textbf{Rouge-2}}                         & \multicolumn{3}{c}{\textbf{Rouge-L}}                         & \multicolumn{3}{c}{\textbf{BERTScore}}                         \\ \cline{2-13}
                       & R           & P           & F1          & R           & P           & F1          & R           & P           & F1          & R           & P           & F1          \\
    \hline
    BART-Base& 53.83          & 56.39          & 53.88          & 36.61          & 39.59          & 37.13          & 51.70          & 54.26          & 51.81          & 74.03  & 74.74 & 74.22\\
    BART-Large& 55.54          & 51.27          & 51.61          & 36.22          & 34.34          & 34.12          & 53.22          & 49.65          & 50.14          & 75.02           & 73.01          & 73.84\\
    ProphetNet& 43.56          & 43.65          & 42.33          & 24.69          & 24.98          & 24.18          & 41.72          & 48.48          & 43.91 & 73.01 & 74.99 & 73.84          \\
    Pegasus& 51.50          & 51.55          & 49.92          & 34.95          & 35.65          & 34.23          & 50.59          & 50.89          & 49.60   & 73.22 & 73.27 & 73.06       \\
    BRIO& 54.09          & 53.55          & 52.58          & 35.23          & 34.82          & 34.06          & 51.37          & 50.96          & 50.00          & 74.04               & 73.32          & 73.52\\
    SimCLS & 55.70 & 45.86 & 48.37 & 33.50 & 26.99 & 28.59 & 52.20 & 45.43 & 47.07 & 68.49 & 72.14 & 70.05 \\
    % SeqCo &           &           &           &           &           &           &           &           &           \\
    \llyr{FOSS-LTE}               & \llyr{13.84}          & \llyr{8.45}          & \llyr{10.25}          & \llyr{0.02}          & \llyr{0.02}          & \llyr{0.02}          & \llyr{10.68}          & \llyr{6.60}          & \llyr{7.97}          &\llyr{44.63}          &\llyr{39.5}          &\llyr{41.87}        \\ 
    \textbf{\tool-Base}& \textbf{56.45} & \textbf{60.54} & \textbf{57.00} & \textbf{42.26} & \textbf{45.94} & \textbf{43.10} & \textbf{56.36} & \textbf{59.38} & \textbf{56.84} & \textbf{76.70} & \textbf{77.88} & \textbf{77.13} \\
    \textbf{\tool-Large} & \textbf{60.04} & \textbf{55.77} & \textbf{56.58} & \textbf{43.45} & \textbf{41.16} & \textbf{41.44} & \textbf{59.02} & \textbf{55.30} & \textbf{56.22} & \textbf{77.37} & \textbf{75.72} & \textbf{76.39} \\
    \bottomrule
    \end{tabular}
    \vspace{-4mm}
\end{table*}

\subsection{RQ2: License Text Summarization} \label{sec:RQ2}

\noindent \textbf{Setup}. \llyr{Since this paper presents the first work towards automated license summarization, there exist no previous works that can be directly used as a baseline. Hence, we carefully select 7 baselines to evaluate the performance of \tool. The motivation for selecting the baselines and why these approaches can be used in the study are as follows. {(1)} The first group of baselines are adapted from existing text summarization techniques. Although license text is usually long and complicated, written in a formal language, it has some common characteristics with natural language. Hence, we utilize three state-of-the-art techniques (i.e., Pegasus~\cite{zhang2020pegasus}, BRIO~\cite{liu2022brio}, and SimCLS~\cite{liu2021simcls}) for abstractive text summarization. Their high performance on various datasets suggest potential capability for summarizing licenses. Like \tool, the input and output of these techniques are original license text and its summary, respectively. We directly applied these techniques as baselines in the experiments. {(2)} The second group of baselines was inspired by the great success of pre-training natural language models across various NLP domains and tasks. In this group, we select three popular and representative pre-training models with acceptable computational cost (i.e., BART-Base, BART-Large~\cite{lewis2020bart}, and ProphetNet~\cite{qi2020prophetnet}). By fine-tuning these models with the license summarization dataset, they have the potential to generate effective summaries for licenses. Note that \tool employs transfer learning and uses a pre-training model as the backbone of the multi-task learning, \tool can be built upon arbitrary language models with the ability to summarize text. The results show that \tool can boost the performance of pre-training models in the task of license summarization. {(3)} Finally, we also used FOSS-LTE~\cite{APSEC`17-Kapitsaki-termsIdentifying}, which was a previous work on license understanding, as a baseline. Although FOSS-LTE was originally proposed for extracting license terms (e.g., \textit{CannotCommercialUse}), we have a similar goal of facilitating license understanding. Therefore, we adapted FOSS-LTE for license summarization by concatenating the regulations extracted by this approach.} 

To conduct a fair comparison, all approaches were first trained on the CNN summarization dataset to obtain the ability of abtractive text summarization, and then fine-tuned on the license summary dataset as described in \Cref{sec:dataset} for the LTS task. The reason behind is that pre-training on CNN helps all baselines and \tool generalize to an unseen, low-resourced, but related domain (i.e., license text in this paper), and thus can boost the performance of all models. 
For each experiment, we conducted a 3-fold cross validation by randomly splitting the entire dataset into three equal parts. In each iteration, one part was designated as the test set while the remaining parts were split into training and validation sets by 3:1. This process was repeated three times so that each part was used as the test set once, and we recorded the average performance of three iterations for comparison.
As previous studies on abstractive text summarization~\cite{lewis2020bart,liu2021simcls,liu2022brio}, we use \textbf{ROUGE-1/2/L}~\cite{lin2004rouge} and \textbf{BERTScore}~\cite{bert-score} as the evaluation metrics.
ROUGE-1/2/L computes the overlap degree between the generated summaries and the referenced summaries in the dataset. Specifically, ROUGE-1 and ROUGE-2 compute the ratio of overlapped unigrams and bigrams to the total number of  unigrams and bigrams in the referenced summary, respectively. ROUGE-L considers the longest common subsequence between the generated and referenced summaries. 
BERTScore %\footnote{Hash: microsoft/deberta-xlarge-mnlk\_L40\_no-idf\_version=0.3.11 (hug\_trans=4.24.0).} 
is an automatic evaluation metric that computes a cosine similarity between each token in a generated summary and that in a labelled summary using contextual embeddings. \lly{Unlike traditional metrics such as ROUGE-1/2/L, BERTScore considers the meaning of words in the context of the sentence, which makes it more effective at measuring semantic similarity.}
The replication package and the implementation details can be seen online~\cite{url-LiSum}.

\noindent \textbf{Result}.
\Cref{expe:summary} shows the results of 7 baselines and \tool for LTS. To conduct a fair comparison, we implemented \tool with two different pre-trained models (i.e., BART-Base and BART-Large), and represented them with \tool-Base and \tool-Large, respectively. Overall, it can be seen that both \tool-Base and \tool-Large outperformed 7 baselines with gains of roughly 5 points on all F1 scores. Specifically, \tool-Base achieves 58.89\%, 45.38\%, 58.98\%, and 78.00\% F1 scores \textit{w.r.t.} Rouge-1, Rouge-2, Rouge-L, and BERTScore, respectively, surpassing the results of all baselines by large margins. 
\llyr{The FOSS-LTE baseline performed significantly worse than the other models, indicating its limited effectiveness in generating accurate and meaningful license summaries. This is expected as FOSS-LTE was not designed specifically for license summarization.}
Comparing \tool-Base and BART-Base, the effectiveness of the proposed multi-task learning can be inferred, since they were constructed based on the same pre-trained model (i.e., BART-Base) and trained on the same datasets (i.e., first CNN and then the referenced summary dataset). \tool-Large achieved a competitive performance with \tool-Base, which also have large improvements on the summarization metrics compared with the baselines. Similarly, based on the same pre-trained model, \tool-Large outperformed BART-Large by at least 5 points of the Rouge scores and 3 points of BERTScore. The results indicate that LTS and LTC are two highly-correlated tasks, and thus the proposed jointly training objective for multi-task learning can improve the performance of license text summarization. 

From \Cref{expe:summary} we can also observe that there exists a relatively large difference between the performance of the baselines \textit{w.r.t.} Rouge-1/2/L. However, in terms of BERTScore, all baselines achieved similar scores which were stably and significantly lower than the scores achieve by \tool-Base and \tool-Large. For example, it can be seen that ProphetNet achieved the worst performance among all 8 models \textit{w.r.t.} Rouge-1/2/L, but its BERTScore was closer to those of other baselines. A possible reason is that the n-gram-based metrics like Rouge-1/2/L rely on string matching, which might fail to match paraphrases in a robust way. In contrast, BERTScore computes the similarity between generated and referenced summaries using contextualized token embeddings~\cite{bert-score}, leading to more robust and semantically-correct results. 

Finally, it can also be seen that models pre-trained on BART-Large (i.e., BART-Large and \tool-Large in \Cref{expe:summary}) were not stably superior to those pre-trained on BART-Base (i.e., BART-Base and \tool-Base). Although BART-Large has exhibit great power in many NLP tasks, the results of BART-Base were close to those of BART-Large in LTS. A possible explanation is that LTS is a low-resourced task, with only 210 licenses for training and testing. Despite being pre-trained on the CNN dataset, the relatively high computational complexity of BART-Large did not bring about significant performance improvement over BART-Base on LTS.

\noindent \fbox{
	\parbox{0.95\linewidth}{
	\textbf{Answer to RQ2:} \tool-Base and \tool-Large outperformed 6 state-of-the-art baselines with gains of at least 5 points \textit{w.r.t.} F1 scores of 4 summarization metrics. The effectiveness of the proposed multi-task learning can be inferred by comparing the performance of the models with the same backbones.}
}

\subsection{RQ3: License Term Classification}\label{sec:RQ3}
\noindent \textbf{Setup}.
Motivated by the user study and TLDRLegal~\cite{url-tldrlegal}, we also provide an extremely short summary of rights and obligations conveyed by a license. We achieved such summaries by a classification task that classified the attitude towards each license term (e.g., \textit{Copy}) into four classes (i.e., \texttt{can}, \texttt{cannot}, \texttt{must}, and \texttt{not mentioned}). Like previous studies~\cite{url-tldrlegal,xu2021lidetector}, we only focused on 23 key license terms as listed in \Cref{tab:terms}. The ground-truth dataset was obtained from TLDRLegal~\cite{url-tldrlegal} and manually validated by the authors. We note that such extremely short summaries may omit some details about the regulations, e.g., ``\textit{changing it is allowed as long as the name is changed}''
will be compressed into \textit{Can modify}. Hence, the classification result only serves as an auxiliary tool for license understanding. 

To evaluate the performance of license term classification (LTC) in the proposed multi-task learning, we compare \tool with 4 popular pre-trained models (i.e., \textbf{BART-Base}~\cite{lewis2020bart}, \textbf{BART-Large}~\cite{lewis2020bart}, \textbf{BERT-Large}~\cite{kenton2019bert}, and \textbf{Roberta-Large}~\cite{liu2019roberta}), which have achieved great success in many natural language processing (NLP) tasks including text classification. Among these baselines, BART-Base and BART-Large are two sequence-to-sequence pre-trained models, and thus can be exploited as the baselines for both LTS and LTC. BERT-Large and Roberta-Large are two state-of-the-art language representation models that can be utilized for classification tasks. Notably, the baselines adopt individual classifiers for each license term, while \tool employs a multi-task learning model with a joint training objective, addressing both LTS and LTC for 23 license terms.
\llyr{In addition to the mentioned baselines, we also include \textbf{FOSS-LTE}~\cite{APSEC`17-Kapitsaki-termsIdentifying} as a comparative baseline. While FOSS-LTE achieves license text term recognition, we modify the attitudes carried by its predefined terms to match our classification task rules (e.g., changing ``may'' to ``can,'', ``no permission/limited'' to ``cannot'').} 
 Our models (i.e., \tool-Base and \tool-Large) in RQ2 and RQ3 are exactly the same models but evaluated \textit{w.r.t.} different aspects. To evaluate the performance of all models, we compute the {recall}, {precision}, {F1 scores}, and calculate the \textbf{micro} and \textbf{macro average} of these scores. The training, validation, and testing sets are the same with those in \Cref{sec:RQ2}.

\begin{table}[]
\footnotesize
    \caption{Results of license term classification (\%)}
    \vspace{-2mm}
    \label{expe:classification}
    \begin{center}
        \begin{tabular}{lcccc}
            \toprule
            \textbf{Model} & \textbf{Micro-F1} & \textbf{Macro-R} & \textbf{Macro-P} & \textbf{Macro-F1} \\
            \midrule
            \llyr{FOSS-LTE} & \llyr{63.83} & \llyr{19.38} & \llyr{24.24} & \llyr{18.80}           \\ 
            \midrule
            BART-Base$^*$          & 85.73 & 32.10 & 30.59 & 29.67  \\
            BART-Large$^*$         & 86.22 & 37.56 & 35.18 & 35.15  \\
            BERT-Large$^*$         & 91.94 & 42.51 & 39.83 & 39.69  \\
            Roberta-Large$^*$      & 88.44 & 37.20 & 34.79 & 34.38  \\
            \textbf{\tool-Base$^*$} & \textbf{93.25}	& \textbf{56.87} &\textbf{52.60} & \textbf{53.56}\\
            \textbf{\tool-Large$^*$} & \textbf{88.62} & \textbf{43.20} & \textbf{39.83} & \textbf{40.13}\\ \midrule %\midrule \midrule
            BART-Base      & 89.49 & 40.44 & 37.16 & 37.42  \\
            BART-Large     & 87.29 & 40.91 & 37.02 & 37.24  \\
            \textbf{\tool-Base}     & \textbf{95.13} & \textbf{62.88} & \textbf{59.82} & \textbf{60.47}        \\
            \textbf{\tool-Large}        & \textbf{91.14} & \textbf{47.03} & \textbf{42.16} & \textbf{43.12} \\
           
            \bottomrule
        \end{tabular}
    \end{center}
        \vspace{-2mm}
        \begin{tablenotes}\footnotesize
            \item \tool-Base and \tool-Large are the same models as those in \Cref{expe:summary}.
            \item Models with $*$ were directly trained on the LTC dataset 
            \item Models without $*$ were first trained on the CNN dataset and then the LTC dataset.
            \item R: recall; P: precision; F1: F1 score.
        \end{tablenotes}
        \vspace{-8mm}
        \iffalse
    \begin{center}
	    \footnotesize
	    {\tool-Base and \tool-Large are the same models as those in \Cref{expe:summary}. \\Models with $*$ were directly trained on LTC, and models without $*$ were \\first trained on CNN and then LTC. R: recall; P: precision; F1: F1 score.}
	\end{center}
 \fi
\end{table}

\noindent \textbf{Result}.
\Cref{expe:classification} shows the results of license term classification for \tool and the baselines.
Note that since each baseline was constructed based on a pre-trained model, we simply represent the baselines with the pre-trained models they used (e.g., BART-Large). However, except multi-task learning like \tool-Base and \tool-Large, the models in \Cref{expe:summary} and \Cref{expe:classification} are different models trained on different datasets for different tasks. Moreover, since BERT-Large and Roberta-Large cannot be directly applied to generate texts, we trained them the license term classification dataset without fine-tuning on CNN. To conduct a fair comparison, we also implemented BART-Base, BART-Large, \tool-Base, and \tool-Large without fine-tuning on the CNN summarization dataset. We use the symbol $*$ to highlight the models without fine-tuning on CNN. We only display the average results for 23 license terms in \Cref{expe:classification}, and the individual results for each license term can be seen online~\cite{url-LiSum}. %We also compare \tool with \tool (only classifier)
Combining the results in \Cref{expe:summary} and \Cref{expe:classification}, it can be seen that the proposed multi-task learning boosted the performance of license text summarization and license term classification simultaneously. The performance improvement brought by jointly training LTS and LTC might come from the high correlation between two tasks. 

As for the classification task, we can observe from \Cref{expe:classification} that with the same training datasets, \tool-Base and \tool-Large outperformed all baselines in terms of four metrics. Among all models and settings, \tool-Base fine-tuned on CNN achieved the best performance with 95.13\% micro average F1, 62.88\% macro average recall, 59.82\% macro average precision, and 60.47\% macro average F1, respectively. The results show that \tool cannot only provide a summary for an arbitrary license, but also a list of license terms with accurate attitudes conveyed by the license. In addition, by comparing the same models with and without $*$ in \Cref{expe:classification}, it can also been inferred that fine-tuning on the CNN dataset benefited both the LTS and LTC tasks. A possible reason is that training on CNN equipped the models with the ability to generate text summaries, which might be highly correlated with license term classification.
\vspace{1mm}

\noindent \fbox{
	\parbox{0.95\linewidth}{
	\textbf{Answer to RQ3:} The proposed multi-task learning boosted the performance of LTS and LTC simultaneously. Among all models, \tool-Base fine-tuned on CNN and the license dataset achieved the best performance with 95.13\% micro average F1, 62.88\% macro average recall, 59.82\% macro average precision, and 60.47\% macro average F1, respectively, surpassing state-of-the-art baselines by large margins.
 }
}

\subsection{Ablation Study}

\begin{table}[]
\footnotesize
    \caption{Ablation study on license representation (\%)}
    \vspace{-1mm}
    \begin{threeparttable}
        \begin{tabular}{lccccc}
            \toprule
               \textbf{License Representation}                    & \textbf{R1} & \textbf{R2} & \textbf{RL} & \textbf{Mi-F1} & \textbf{Ma-F1}   \\
            \midrule
            \tool-Base (encoder) & 56.88 & 42.59 & 56.75 & 93.94 & 54.13\\
            \tool-Base (decoder) & 54.32 & 39.69 & 54.10 & 66.01 & 19.03 \\
            \tool-Base (concat)  & \textbf{57.00} & \textbf{43.10} & \textbf{56.84} & \textbf{95.13} & \textbf{60.47}\\ \midrule
            \tool-Large (encoder) & 52.46 & 35.63 & 51.28 & 85.40 & 32.60 \\
            \tool-Large (decoder) & 52.78 & 36.75 & 52.17 & 78.49 & 27.43 \\
            \tool-Large (concat) & \textbf{56.58} & \textbf{41.44} & \textbf{56.22} & \textbf{91.14} & \textbf{43.12}\\
            % \tool(FIXED)   & Very Poor \\
            
            \bottomrule
        \end{tabular}
        \begin{tablenotes}\footnotesize
            \item R1: ROUGE-1; R2: ROUGE-2; RL: ROUGE-L; Mi: micro; Ma: macro.
            \item All the results in the table are F1 scores.
        \end{tablenotes}
    \end{threeparttable}
    \vspace{-3mm}
    \label{tab:encoder}
\end{table}

Given a license text, \tool represents it by concatenating the last hidden states of the encoder and the decoder, and feeds it into the license term classifier for multi-task learning. To investigate the effects of different ways of license representation, we varied the representations of licenses and controlled the other modules of \tool unchanged.
Specifically, we represented the licenses with the last hidden states of the encoder and the decoder, and separately fed them into the classifier for jointly training.
\Cref{tab:encoder} displays the results of the ablation study on license representation. We use encoder, decoder, and concat to denote license represented by the outputs of the encoder, decoder, and the concatenation of them, respectively. It can be seen that by concatenating the outputs of the encoder and decoder, \tool achieved the best performance in both summarization and classification. In addition, while \tool (decoder) achieved competitive performance with \tool (encoder) in the summarization metrics, its performance in classification was much worse than \tool (encoder). The performance degradation might be because of the information loss in the decoder outputs. For the reason that the decoder was trained to output license summaries, compared with the encoder of licenses, some information might be lost, resulting in a poor classification result. 
\section{Discussions}
{\noindent\textbf{Insights and Lessons Learnt.} \llyr{{(1)} The comparison between models trained on a single task (i.e., license text summarization or license term classification) and \tool which leverages multi-task learning shows that the two tasks are highly correlated and thus jointly learning these tasks can boost their performance simultaneously. {(2)} The performance improvement from \tool-Base to \tool-Large is not as substantial as from BART-Base to \tool-Base. This suggests that while scaling up the model can lead to performance gains, there might be diminishing returns with larger models for this particular task, possibly due to the limited amount of data. {(3)} The results reveal a notable discrepancy between the high Micro-F1 and relatively lower Macro-F1 scores in license classification for all models, indicating the presence of class imbalance in the dataset. This imbalance can lead to biased model predictions and affect the performance on minority classes. To address this issue, we plan to present balancing techniques to enhance the performance for all license terms in the future work.}}

\noindent \textbf{Threats to Validity.}
(1) \tool utilized the HuggingFace implementations of the pre-trained models~\cite{url-huggingface} as the backbones. For this reason, the effectiveness of the HuggingFace implementations might affect the accuracy of \tool.
(2) \tool was constructed based on BART, a widely-used and popular sequence-to-sequence pre-trained model. However, the effectiveness of the backbone might also affect the performance of \tool. Thus, one of the future plans is to exploit other pre-trained models to investigate their effects on the performance of \tool. 
(3) License text is a low-resourced language. As a result, the sizes of the test sets are not large, which might bring threats to the validity of experiments. To mitigate this problem, we employed a three-fold cross-validation to select hyper-parameters and computed the average scores achieved by \tool.
(4) We conducted a user study to investigate the perspectives and practices of the participants towards OSS licenses. %Specifically, we recruited 661 participants via two online research platforms. 
However, the participants might not be representative enough. To address this issue, we recruited 661 participants across different countries and performed purposive sampling to conduct the survey. 

\noindent \textbf{Limitations.}
(1) \tool exploited BART as the backbone for license understanding. However, when license texts are too long, both BART-Base and BART-Large truncate the input texts, hence, some information might be lost in this case. (2) As previous studies~\cite{url-tldrlegal, xu2021lidetector}, \tool only inferred the attitudes towards 23 key license terms. There might exist some rights or obligations omitted by the license term classification provided by \tool. 
\section{Related Work}
\subsection{License Comprehension}
To facilitate license understanding, much research has been done to help developers comprehend license texts.

\noindent \textbf{The ontology study}. A majority of studies on license understanding conducted the ontology study on licenses. Specifically, Alspaugh et al.~\cite{IREC`09-Alspaugh-Intellectual,AIS`10-Alspaugh-Challenge} modeled 10 licenses by extracting a set of tuples (e.g., actors, actions, and objects) from license texts. 
Gordon et al.~\cite{Qualipso`10-gordon-prototype} exploited the Web Ontology Language (OWL) to manually construct the ontology that incorporate knowledge from 8 popular licenses. Based on the ontology, they analyzed license compatibility issues for OSS projects~\cite{ICAIL`11-Gordon-Analyzing}. They implemented a license analyzer named MARKOS~\cite{ICAIL`13-Gordon-Introducing,COMMA`14-Gordon-Demonstration} for license understanding and incompatibility analysis. Despite the efficiency, the ontology-based studies relied on manually extracting tuples from license texts, and the information was extracted only from a small set of popular licenses, which may not be suitable and flexible when applied for other licenses. 

\noindent \textbf{Statistical learning}. To automatically extract the regulations stated by arbitrary licenses, FOSS-LTE~\cite{APSEC`17-Kapitsaki-termsIdentifying} presented the first license understanding tool based on a topic model. It manually analyzed 25 licenses, summarized the phrases related to license terms, and classified license terms into rights, obligations, and conditions. Then, it automatically identified license terms from license texts by first mapping licenses terms with a set of predefined phrases, and then mapping phrases with topics by Latent Dirichlet Allocation (LDA). %In this way, it mapped license terms with topics for automated extraction of license terms. 
Although FOSS-LTE can automatically extract license terms from license texts, it relied on manual analysis to map license terms with topics, and the topic model may also introduce much noise, leading to a low accuracy for license term extraction. 

\noindent \textbf{Machine learning}. To overcome the aforementioned issue, Xu et al.~\cite{xu2021lidetector} proposed a machine learning-based method for license understanding and incompatibility analysis. They first employed sequence labelling to extract license terms from license texts, which is capable of extracting flexible expressions for license terms. 
Then they performed grammar parsing and sentiment analysis to infer the attitudes conveyed by the copyright holder given the extracted license terms. By this means, they automatically extracted the rights and obligations from license texts.
Unlike previous studies which focused on license term extraction, in this paper, we conduct the first study on license text summarization, aiming at providing users a breif and precise summary to facilitate license understanding.

\subsection{Text Summarization}
There are mainly two types of text summarization approaches, i.e., extractive and abstractive. 

\noindent \textbf{Extractive text summarization}. This type of summarization can be regarded as a binary classification task~\cite{liu2019text}, which labels each token in the input document to indicate whether the token is selected or not into the summary. SummaRuNNer~\cite{nallapati2017summarunner} presented a sequence model based on the Recurrent Neural Network for extractive test summarization. Chen et al.~\cite{chen2018iterative} improved the understanding ability of extractive summarization by making the model repeatedly polish the distributed representation of the document.
Although extractive methods have achieved high performance~\cite{lin2004rouge}, they often suffered from the problem of redundancy and low readability, and the generated summaries might contain conflicting information~\cite{hsu2018unified}.

\noindent \textbf{Abstractive text summarization}. Unlike extractive text summarization, abstractive text summarization understands the main concept of the documents and generates new words to compress the document. 
BART~\cite{lewis2020bart} is a widely-used sequence-to-sequence model, and has proved its effectiveness in many NLP tasks especially for text generation.
ProphetNet~\cite{qi2020prophetnet} and Pegasus~\cite{zhang2020pegasus} are two pre-trained sequence-to-sequence models with self-supervised pre-training objectives.
To improve the summarization performance, SimCLS~\cite{liu2021simcls} first generated candidate abstracts and then selected the best candidate evaluated with another pre-trained RoBERTa~\cite{liu2019roberta}. 
Similarly, BRIO~\cite{liu2022brio} ranked the generated candidate sequences and indirectly incorporated the ranking process into the optimization objective. It has achieved state-of-the-art performance on CNN/DailyMail~\cite{hermann2015teaching} and XSum~\cite{narayan2018don}.
\llyr{However, directly applying traditional text summarization techniques for license text summarization may yield subpar results. This can be mainly attributed to the overlooking of license texts as mere textual content, failing to recognize their emphasis on rights and obligations. To address this challenge, we propose a multi-task learning algorithm which leverages the correlation between licnese text summarization and license term classification to incorporate more information and make the model focus on the rights and obligations stated by licenses.}

\section{Conclusion}
In this paper, we first conducted a user study to explore the perspectives and practices of developers towards OSS licenses. The user study revealed that 71.43\% of the participants have encountered the difficulty of choosing OSS licenses. 69.29\% of the participants thought OSS licenses were too long and complicated. Motivated by the user study and the fast growth of licenses, we propose the first study towards automated license summarization. Specifically, we present \tool, a multi-task learning method to help developers overcome the obstacles of understanding OSS licenses. Comprehensive experiments demonstrated that the proposed jointly training objective boosted the performance on both tasks, surpassing state-of-the-art baselines with gains of at least 5 points \textit{w.r.t.} F1 scores of four summarization metrics and achieving 95.13\% micro average F1 score for classification simultaneously. Moreover, the results of the ablation studies demonstrated the effectiveness of each module in \tool. We released all the datasets, the replication package, and the questionnaires to facilitate follow-up research.
\section*{Acknowledgements}
This work was supported by the National Key Project of China (No.2020YFB1005700) and the National Natural Science Foundation of China (No.62202245 and 62002178).

\clearpage
\bibliographystyle{IEEEtran}
\bibliography{IEEEabrv,paper-ref}

\clearpage

\end{document}